\documentclass[letterpaper]{article}

\usepackage{natbib,alifeconf}  

\title{An Ising-like Model for Language Evolution}
\author{Conor Houghton \\
\mbox{}\\
Faculty of Engineering, University of Bristol, UK\\
conor.houghton@bristol.ac.uk} 

\begin{document}
\maketitle

\begin{abstract}
I propose a novel Ising-like model of language evolution. In a simple way, Ising-like models represent the countervailing tendencies towards convergence and change present in language evolution. In the ordinary Ising-model, a node on a graph, in this case representing a language speaker, interacts with all its neighbors. In contrast, in the model proposed here, a node only interacts with the neighboring node whose state-vector is most similar to its own. This reflects the tendency of people to interact with others who speak a similar language. Unlike the ordinary Ising model, which tends towards language continua, this new model allows language boundaries. 
\end{abstract}

Languages evolve under the influence of contrary forces, forces that encourage convergence and those that encourage change. For a start, languages are only useful insofar as they are understood. Under this imperative an individual's language should align with the languages of others. However, there is a contrary propensity towards language invention: an inclination, particularly among the young, to modify or reinvent language, either to exclude other, perhaps, older speakers or out of a simple delight in the act of language creation. Another cause of change is found in grammar. Here a move towards a more explicit and logical grammar, one that aids the speaker and listener in the precise use of language, is opposed by a sort of laziness, a desire, even at the cost of inconsistency and potential ambiguity, to employ shorter or sloppier language or to find habitual short-hand forms for frequently used expressions. 

There have been useful and informative attempts to model language evolution. For example the iterated language models simulate the emergence of compositionality in languages \citep{KirbyHurford2002,SainsHoughtonBullock2023}. However, my goal here is to suggest the simplest possible model that encompasses the processes of convergence and change outlined above. This leads me to a simple Ising-like model I will call ``the preference model''. In this extended abstract, I will describe the model and my motivation in proposing it. A more detailed comparison of the properties of language evolution in this model is not attempted, but I believe that this would be interesting. I also believe that this model is of interest in-and-of itself and that it could be extended to include other forces that shape language, such as the tendency towards internal consistency.

In an Ising model the nodes of a graph have a value of plus or minus one; this value is called the \textsl{spin} of the node. The Ising model is used in physics to model magnetization and is important because it has a phase transition and because, in particular cases, it is a solvable thermodynamic model \citep{Onsager1944}. In physics the overall energy of the system is important and this energy is minimized by aligning the spins of a node with those of its neighbors in the graph.
It is, however, a thermodynamic model and so the nodes do not always change their states in a way that reduces this energy. The ``Metropolis'' formation will be used here. At each time step a node is chosen and the consequence of changing its value, from plus to minus, or minus to plus, is investigated. If the current spin of the node $x$ is $s_x$ then the change to the energy that would result from flipping the sign of $s_x$ is
\begin{equation}
dE=\frac{2}{n}s_x\sum_y s_y
\end{equation}
where the sum is over all connected nodes and $n$ is the degree of the nodes. If $dE$ is negative, then flipping $s_x$ lowers the energy and this change is accepted. If it is positive it is accepted with probability
\begin{equation}
    p=\exp{(-dE/T)}
\end{equation}
where $T$, the temperature, determines the magnitude of the random thermal effects. 

\begin{figure*}[ht]
\begin{center}
\begin{tabular}{lll}
\textbf{A}&\textbf{B} ordinary&\textbf{C} preference\\
\includegraphics[height=1.5in]{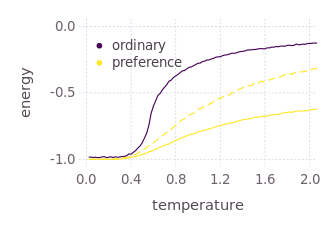}&\includegraphics[height=1.5in]{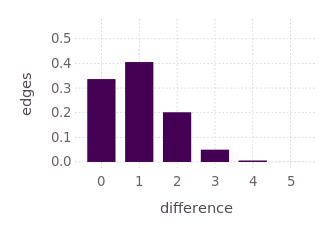}&\includegraphics[height=1.5in]{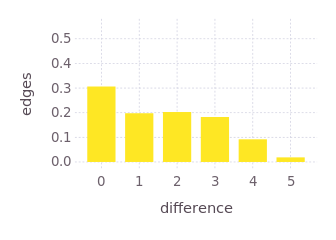}
\end{tabular}
\caption{\textbf{Comparing the ordinary and preference model}. The solid lines in \textbf{A} plot the energy as a function of temperature. Finite size effects smooth the phase transition in the ordinary model, it is nonetheless visible at around $T=0.57$. For the preference model, there is a more gradual change in behavior at the same temperature; since the energy in the preference model is calculated between a node and its most similar neighbor the $T\rightarrow \infty$ limit is not zero. In the dashed line the energy is ``centered'' and normalized so that it takes values from -1 to zero. \textbf{B} and \textbf{C} plot the similarity of nodes on each side of an edge. For each edge, the Hamming difference between the states is histogrammed and changed to a probability. In the preference model, neighboring nodes differ considerably more. For all three plots $L=5$, the grid is $50\times 50$; the average result of ten trials is shown and for each trial the model is run for 25,000 time steps to reach equilibrium. In \textbf{A} the energy is the average for each spin-spin connection. In \textbf{B} and \textbf{C} $T=0.3$.}
\label{figure}
\end{center}
\end{figure*}

In this way, the Ising model models the key aspect of language evolution noted above. There is a competition between alignment and randomness and so the simplest putative Ising-like model of language evolution would simply be an Ising model on a two-dimensional square lattice: the lattice representing the geographical distribution of speakers. However, this would only allow for two languages, the ``up-language'' and the ``down-language''. To address this shortcoming, the individual spin $s_x$ is replaced with a length $L$ vector of spins $\textbf{s}_x$ that will be referred to as the state of the node. The idea is that each of the individual spins corresponds to a property of the language. Thus, for example one spin in $\textbf{s}_x$ might be thought of as determining the order of noun and adjective. In this model, a node is chosen at random and a component of that node's state is selected, again randomly. This component is flipped, or left unflipped, using the same thermodynamics described above. If the state has length $L$ this model of language evolution is, effectively, $L$ independent Ising models. It would be possible to compare this model to the distribution of languages, attempting to use the distribution of cluster sizes for example, to fix a value of $T$ and $L$. Something like this is done using a different model in \cite{SivaEtAl2015, SivaEtAl2017} with interesting results.

However, there is a problem with this model. Famously, a putative nineteenth century traveler could walk from Lisbon to Naples without crossing a language boundary. Although Portuguese and Neapolitan are very different languages, people living near each other were always able to communicate. This is a property of the $L$-state Ising model. However, in the real world, language continua are common but not universal. If the putative traveler varied their route just a small bit they would pass through the Basque country and would certainly cross a language boundary. To explain this, I note that an infant does not poll its neighbors and use a language mixture as an exemplar. Rather it typically learns from the people in its household, often parents, and after that will preferentially communicate with other people who speak a similar language to theirs. Here, I am proposing that this be incorporated into our simple model of language evolution, somewhat in the spirit of the bounded confidence model \citep{HegselmannKrause2019}: the model is modified that nodes only align with preferred neighbours.

Here, I call this the \textsl{preferential model} and the original $L$-state Ising model, the \textsl{ordinary model}. In the preferential model, at each time step, a node and spin are chosen at random as before. Next, however, the Hamming distance is calculated between the state of the node and the states of its four neighbors. The closest of these is selected. This might involve a random selection if there are equally close nodes. The same thermodynamical calculation, made using only the node and its closest neighbor, then decides whether or not to flip the spin. This solves the problem: in the preferential model there is a mixture of language continua and language boundaries. This is illustrated in Fig.~\ref{figure}. 

I suggest that the preferential model is an interesting, simple model of language evolution. There are lots of potential variations of the model, such as local temperature changes, weighted random selection of the preferred neighbor and interaction between components of the state vector. However, before consider further variants, the properties of the current model will have to be studied. What is the nature, for example, of the transition around the temperature where the ordinary model has its phase transition? The model also needs to be compared to language data to decide if this simple model has any potential to describe, in a meaningful and interesting way, the distribution of languages. 

\section{Code availability}
\texttt{github.com/evoising/alife2023}.
\section{Acknowledgments}
Thanks for Jake Witter who did computer simulations on an early XY-model version of the model presented here. I am a Leverhulme Research Fellow (RF-2021-533).
\newpage
\footnotesize
\section{Appendix}
Since this is an extended abstract I did not include the formula for energy in the version that appears in ALIFE2023. To avoid confusion with the choice of constants, this is provided here. The energy for the system in Fig.~\ref{figure} corresponding to the ordinary model is
\begin{equation}    
    E=-\frac{1}{nM^2L}\sum_{x\in \mathcal{G}}\sum_{y\in \mathcal{N}(x)} \textbf{s}_x\cdot\textbf{s}_y    
\end{equation}
where $M=50$ is the side-length of the square grid, $L=5$ is the length of the vector representing state and $n=4$ is the degree of the nodes; $\mathcal{G}$ is the set of all nodes and $\mathcal{N}(x)$ is the set of $n$-neighbours to node $x$. Including $n$ in the normalization is convenient for comparing the ordinary and preference models. Relating this to the usual statement of the model, the connection strength is $J=0.25$ rather than one giving a critical temperature of $T\approx 0.57$ rather than the usual $T\approx 2.27$. 

It should also be noted that this describes $L$ independent $M\times M$ Ising models, not an $L\times M\times M$ Ising model. There is no interaction between spins in an individual state. The spins are coupled in the preference model, but in a more complicated way. Including a direct interaction between the spins would be a way to include the tendency of languages towards consistency, the relationship between the order of verb and object and the order of adjective and noun, for example. However, this is not considered here. 

For the preference model
\begin{equation}    
    E=-\frac{1}{nM^2L}\sum_{x\in \mathcal{G}}\textbf{s}_x\cdot\textbf{s}_{y_*}    
\end{equation}
where, now, $n=1$ and $y_*$ is the $y\in\mathcal{N}(x)$ with the largest value of $\textbf{s}_x\cdot\textbf{s}_y$, or, equivalently in this case, the smallest Hamming distance. As it stands, the preference model is described entirely in terms of the update rule; this aspect of the preference model is certainly less elegant than the ordinary model in which the update rule is one of a number of equivalent routes a model described in terms of the distribution of configurations

For high temperatures, spins are more-or-less random. As such, for the ordinary model the energy approaches zero as $T$ increases. However, for the preference model, because the preference is for the closest node, the asymptotic value of the energy is not zero. It is likely the asymptotic value could be calculated analytically. Here, I calculated it numerically and found a value of $E_a\approx -0.448$; the dashed line in Fig.~\ref{figure}\textbf{A} is $(E-E_a)/(1+E_a)$. This is convenient since it allows an easy comparison between the ordinary and preference models and between the preference model with different values of $L$. In fact, the behavior is remarkably unchanged as $L$ is changed: 
\begin{center}
\includegraphics[height=1.5in]{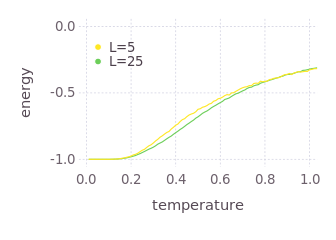}
\end{center}
where, here the $L=5$ and $L=25$ models are compared. For $L=25$ the asymptotic value is $E_a\approx -0.204$ and $E_a$ does get closer to zero as $L$ increases because the distribution of Hamming distances between random state-vectors will be more tightly distributed around its mean. The histogram of node similarities, the equivalent of Fig.~\ref{figure}\textbf{B}/\textbf{C} is
\begin{center}
\includegraphics[height=1.5in]{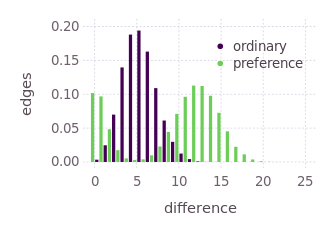}
\end{center}
The ordinary model is 25 independent Ising models and so chance leave few adjacent nodes identical. In the preference model, the models are coupled and the distribution has more identical adjacent nodes and more that are more different. Even in the preference model the number of identical neighboring nodes is small. Adjusting the temperature appears to affect how similar ``preferred'' nodes are
\begin{center}
\includegraphics[height=1.5in]{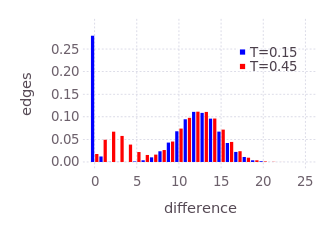}
\end{center}
where, if comparing this graph to the previous one, the difference in
the $y$-scale should be noted. Of course these distributions depend on
the neighbourhood structure; the this-one-not-that-one choice of one
preferred neighbour out of four, a fuller exploration of the model
would include other structures.

\bibliographystyle{apalike} \bibliography{2023_EvoIsing_bib}

\end{document}